\documentclass[%
 reprint,
superscriptaddress,
 amsmath,amssymb,
 aps,
]{revtex4-1}

\usepackage{graphicx}
\usepackage{dcolumn}
\usepackage{bm}
\usepackage{multirow}
\usepackage{amssymb}
\usepackage{amsmath}
\usepackage{epsfig}
\usepackage{bm,multirow}
\usepackage{url}
\usepackage{breakurl}   
\usepackage{lineno}

\usepackage{todonotes}

\begin{document}

\title[Structure of Pitch-Pattern Motifs in Major League Baseball]{Structure of Pitch-Pattern Motifs in Major League Baseball}

\author{Youngjai \surname{Park}} 
\affiliation{Department of Applied Physics, Hanyang University, Ansan 15588, Korea}

\author{Cheawoon \surname{Lim}}
\affiliation{Department of Applied Physics, Hanyang University, Ansan 15588, Korea}

\author{Seung-Woo \surname{Son}}
\email{sonswoo@hanyang.ac.kr}
\affiliation{Department of Applied Physics, Hanyang University, Ansan 15588, Korea}
\affiliation{Center for Bionano Intelligence Education and Research, Hanyang University, Ansan, 15588, Korea}

\author{Mi Jin \surname{Lee}}
\email{mijinlee@pusan.ac.kr}
\affiliation{Department of Physics, Pusan National University, Busan 46241, Korea}
 
\date{\today}


\begin{abstract}
Baseball consists of two teams alternating between batting and fielding while competing to score runs through sequential pitching events. Recent advances in tracking technology have enabled all Major League Baseball (MLB) clubs to record every pitch with high resolution, yet most quantitative studies have primarily emphasized single-pitch metrics, leaving the role of sequential structure less explored. Here, we examine pitch-pattern motifs of multiple lengths using approximately 12.4 million Statcast pitch recordings from the 2008--2025 MLB regular seasons at two complementary scales. At the macroscale, we quantify pitch-sequence diversity using the Shannon entropy and inverse Simpson index and examine their relationships with earned run average and wins. At the microscale, we compare hit and out frequencies across pitch-pattern motifs. Rather than identifying outcome-determining sequences, we find that motif usage exhibits stable, non-random organization, as reflected in Zipf's and Heaps' laws, while showing limited association with conventional performance measures. While language-like scaling (Zipf's and Heaps' laws) clearly reveals an underlying `grammar' of MLB pitch sequences, that grammar alone is insufficient to account for performance indicators such as earned run average or wins. These results suggest that sequence-based analyses clarify the structural organization of pitch usage, while also delineating the limits of motif-based approaches for explaining performance without richer contextual information.

\end{abstract}


\maketitle

\section{Introduction}
\label{sec:intro}
Baseball is one of the oldest and most popular bat-and-ball games, in which two teams alternate between offense and defense while competing to score runs. A standard defensive alignment consists of nine players positioned at pitcher, catcher, first baseman, second baseman, third baseman, shortstop, left fielder, center fielder, and right fielder~\cite{mlb_dot_com,wiki-glossary}. On offense, batters attempt to put the ball in play, allowing runners to advance counter-clockwise across four bases (first, second, third, and home) to score runs. Each game is structured into nine innings, with both teams taking turns batting and fielding in each inning, and the team with more runs at the end of the game is declared the winner. To relieve pitchers of offensive responsibilities and strengthen lineups, the designated hitter rule was introduced and has been adopted league-wide in Major League Baseball (MLB) since the 2022 season.

\begin{table}[b]
\caption{\textbf{Overview of the MLB pitching dataset.} ``Raw data'' refers to the original dataset scraped from the website, without any preprocessing. ``Filtered data'' include only pitches for which all of the following information is available: the $x$ and $z$ locations at home plate, pitch type, and release speed. ``Qualified data'' consists of records from qualified pitchers who threw at least 162 innings in a season, excluding the 2020 season which was shortened to 60 games due to the COVID-19 pandemic. $N_{\rm{pitcher}}$ indicates the number of unique pitchers in the dataset. $N_{\rm{(pitcher,\,year)}}$ means the number of pitcher, year combinations. $N_{\rm{pitch}}$ shows the total number of pitches in the dataset.}
\label{table:overview}
\begin{ruledtabular}
\begin{tabular}{rrrr}
 Stat           & $N_{\rm pitcher}$ & $N_{\rm (pitcher,\,year)}$ & $N_{\rm pitch}$ \\ \hline
 Raw data       & 3,468         & 13,613               & 12,403,853  \\
 Filtered data  & 3,468         & 13,611               & 12,322,641  \\
 Qualified data &   393         &  1,174               &  3,542,338  \\
\end{tabular}
\end{ruledtabular}
\end{table}

\begin{figure*}[t]
\includegraphics[width=\linewidth]{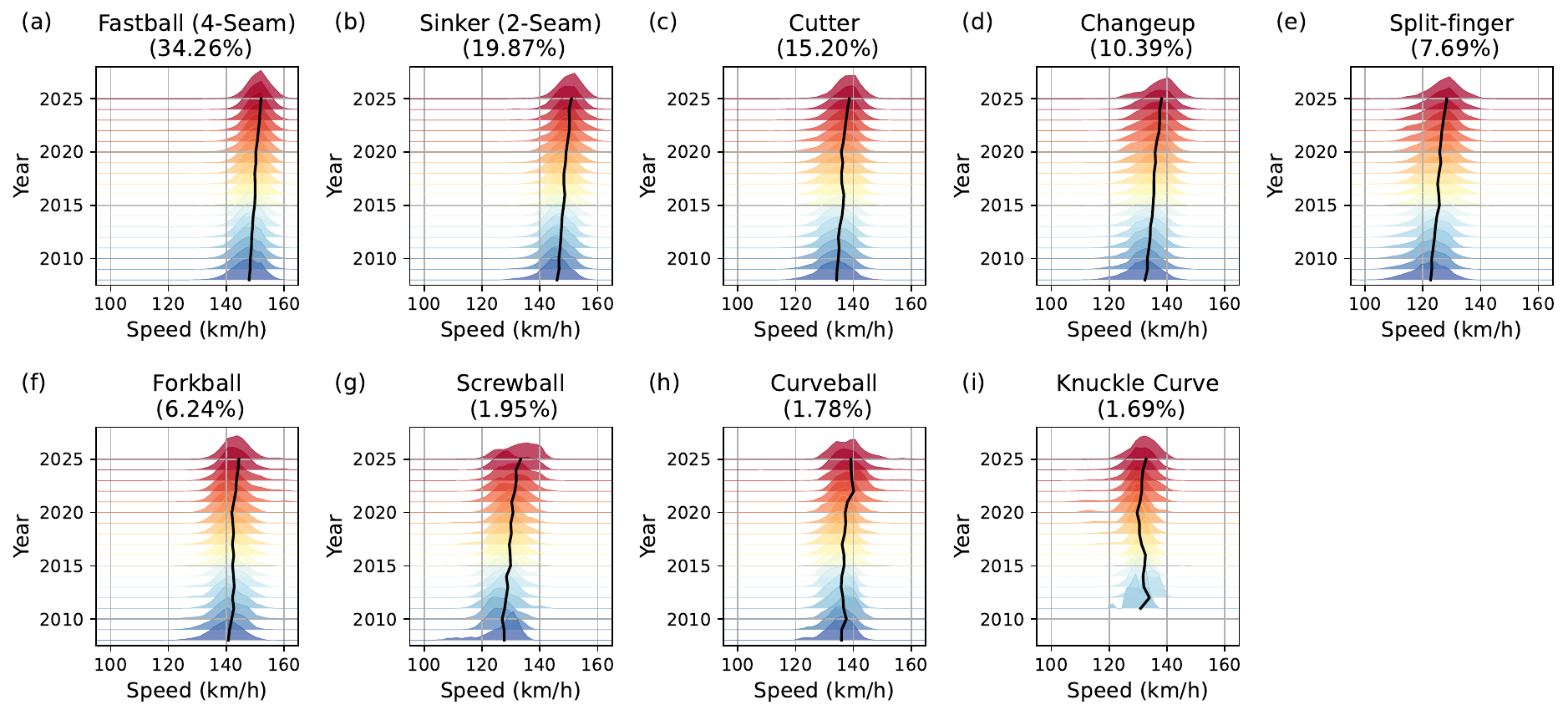}
\caption{\textbf{Pitch-speed distribution over the season.} In the MLB dataset, pitch types are classified into 18 categories~(see Table~\ref{table:pitch_type}). We show the seasonal trends in pitch release speed, observed from the 2008 to 2025 seasons, excluding pitch types that account for less than 1\% of all pitches: (a) fastball, (b) sinker, (c) cutter, (d) changeup, (e) split-finger, (f) forkball, (g) screwball, (h) curveball, and (i) knuckle curve. In each panel, we visualize the annual distribution of release speed using smooth density shapes for each year, and indicate the seasonal average with the black line. Overall, most pitch types exhibit a gradual increase in release speed across seasons. All statistics use the filtered dataset containing 12,322,641 pitches.}
\label{fig:speed_trend}
\end{figure*}

Within this structured and sequential nature of baseball games, quantitative analysis has long played a central role in understanding and improving team performance~\cite{vazquez2007most,jeong2012traveling}. For decades, baseball analysts have leaned on mathematical and statistical tools collectively known as \emph{sabermetrics} to improve team performance~\cite{rosner1996modeling,puerzer2002scientific,seung2012study,hamilton2014applying,healey2015modeling,lee2016suggestion,kwon2023analysis}. A representative example is ``Moneyball,'' which popularized the use of sabermetrics for roster construction and lineup optimization~\cite{lewis2004moneyball}. Yet these approaches have largely emphasized hitters, focusing on reordering batting lineups or valuing on-base skills to boost offensive output. Preventing runs is equally crucial, but previous quantitative analysis is often confined into single-pitch attributes or batter outcomes~\cite{sokol2003robust,jensen2009hierarchical,mcshane2011hierarchical,kwon2023analysis}, leaving the complexity of pitching patterns underexplored. Because pitchers become harder to predict when they vary their pitch usage, examining the diversity and structure of pitch sequences may offer additional insight into how runs can be effectively suppressed~\cite{lee2025pitcher}.

Pitch-pattern analysis faces a fundamental challenge in that pitch labels depend strongly on a pitcher’s subjective judgment: pitches with nearly identical release speeds and trajectories may be tagged differently depending on count, game plan, or individual preference. Such subjectivity hinders consistent comparisons across pitchers and seasons. To address this issue, we compress the 18 Statcast pitch types, defined by measured release speed, movement, and trajectory, into six representative groups with aligned physical characteristics~\cite{Statcast}. In order to explore sequential patterns of pitches, we analyze length-$L$ sequences of pitch groups in the temporal records at the group level, which we define as length-$L$ motifs, rather than individual pitch types. This extends the concepts of motifs as structural building blocks from network subgraphs~\cite{network_motif} to time-ordered pitch sequences, without explicitly considering network connectivity in this study.

Temporal motif analysis based on coarse-grained pitches (pitch groups) provides a compact representation of pitch sequences that captures sequential correlation structures beyond what single-pitch statistics can reveal~\cite{park2021motif}. The group-based method avoids the combinatorial explosion that arises when analyzing individual pitch labels~\cite{motif_complexity}, while still enabling consistent comparisons across pitchers and seasons. Such motif statistics highlight recurrent temporal patterns in pitch usage, offering a potential basis for strategic interpretation of pitching behavior, for example, by identifying significantly overused sequences and motivating deliberate variation to reduce predictability in subsequent games.

Using MLB regular-season Statcast data from 2008--2025~\cite{mlb_dot_com,Statcast}, we construct group-based pitch sequences for each pitcher and each season and decompose them into motifs of various lengths $L\in\{1,2,3,4,5\}$. To assess both the functional relevance and the structural organization of these pitch sequences, we address three questions: (i) how motif diversity (Shannon entropy and inverse Simpson index) relates to conventional performance indicators such as ERA and wins; (ii) whether specific motifs are associated with HIT or OUT outcomes; and (iii) whether motif usage exhibits language-like structural scaling, as characterized by Zipf's and Heaps' laws.

The remainder of this paper is organized as follows. Section~\ref{sec:method} details the pitch reclassification scheme and the motif-based representation. Section~\ref{sec:results} presents the resulting diversity metrics and examines their relationships with pitcher outcomes. Section~\ref{sec:discusssion} discusses broader implications and limitations of the analysis, and Appendix~\ref{sec-app:pattern} provides supplemental dataset information and additional motif-level statistics.

\section{Overview of Major League Baseball Dataset}
\label{sec:data}

\begin{table*}[t]
\caption{\textbf{MLB Teams by league and division.} MLB consists of two primary leagues: the National League (NL) and the American League (AL). Each league is further divided into three divisions (East, Central, and West). Every division is composed of five teams, making for a total of 30 MLB teams. The table below illustrates how these teams are aligned according to their respective league and division.}
\label{table:mlb_team}
\begin{ruledtabular}
\begin{tabular}{|ccc|}
\multicolumn{3}{|c|}{\textbf{National League (NL)}} \\ \hline
\textbf{East} & \textbf{Central} & \textbf{West}    \\ \hline \hline
Atlanta Braves (ATL)       & Chicago Cubs (CHC)       & Arizona Diamondbacks (ARI)\\
Miami Marlins (MIA)        & Cincinnati Reds (CIN)    & Colorado Rockies (COL)    \\
New York Mets (NYM)        & Milwaukee Brewers (MIL)  & Los Angeles Dodgers (LAD) \\
Philadelphia Phillies (PHI)& Pittsburgh Pirates (PIT) & San Diego Padres (SD)     \\
Washington Nationals (WSH) & St. Louis Cardinals (STL)& San Francisco Giants (SF) \\
\hline \hline \hline
\multicolumn{3}{|c|}{\textbf{American League (AL)}} \\ \hline
\textbf{East} & \textbf{Central} & \textbf{West}    \\ \hline \hline
Baltimore Orioles (BAL) & Chicago White Sox (CWS)   & Houston Astros (HOU)     \\
Boston Red Sox (BOS)    & Cleveland Guardians (CLE) & Los Angeles Angels (LAA) \\
New York Yankees (NYY)  & Detroit Tigers (DET)      & Oakland Athletics (OAK)  \\
Tampa Bay Rays (TB)     & Kansas City Royals (KC)   & Seattle Mariners (SEA)   \\
Toronto Blue Jays (TOR) & Minnesota Twins (MIN)     & Texas Rangers (TEX)      \\
\end{tabular}
\end{ruledtabular}
\end{table*}

To analyze pitch-pattern motifs, we combine two Major League Baseball (MLB) data resources: (i) \textit{Baseball Savant}, whose \textit{Statcast Search} interface provides pitch-by-pitch tracking data for every MLB game since the 2008 year~\cite{Statcast}; and (ii) \textit{the Official Site of Major League Baseball} (\texttt{MLB.com}), which supplies historical player statistics dating back to the 1876 year~\cite{mlb_dot_com}. Together, these complementary resources allow us to integrate high-resolution pitch-level measurements with pitcher-level historical context.

From 2008 onward, the raw datasets contain 12,403,853 pitch records thrown by 3,468 pitchers (see Table~\ref{table:overview}). After removing records with missing information (e.g., pitch types or release speeds), 12,322,641 pitches from 3,468 pitchers remain available for analysis. Each Statcast record provides the pitch’s $x$ and $z$ coordinates as it crosses home plate, along with its pitch type and release speed. Pitch speeds increase overall across all pitch types over time (see Fig.~\ref{fig:speed_trend}), a trend that is broadly consistent with the growing influence of modern biomechanics and sports science~\cite{massgeneralbrigham_biomechanics,stodden2005relationship,makhni2018assessment}.

\begin{figure}[b]
\includegraphics[width=\linewidth]{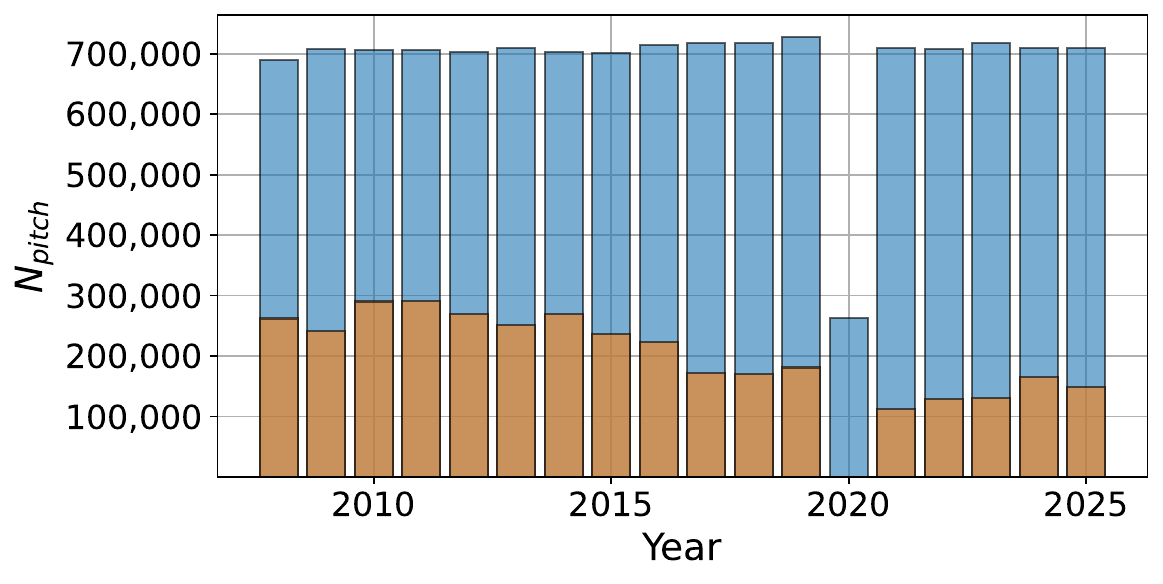}
\caption{\textbf{Total number of pitches across seasons.} The combined height of the blue and orange bars indicates the total number of pitches per season, while the orange bars alone represent the number of pitches thrown by qualified pitchers. Except for the shortened 2020 season due to the COVID-19, the total pitch count remains around 700,000 per year. Notably, despite the relatively stable overall pitch volume, the proportion contributed by qualified pitchers has steadily declined in recent years.
}
\label{fig:n_pitch_year}
\end{figure}

\begin{table}[b]
\caption{\textbf{Statcast-defined six coarse-grained pitch groups by release speeds.} The Statcast classifies pitches into 18 distinct types based on their physical characteristics, such as movement and speed, using their internal classification criteria~\cite{Statcast}. These 18 pitch types are grouped according to their release speeds, yielding six coarse-grained groups. The table lists each pitch type along with its corresponding abbreviation.}
\label{table:pitch_type}
\begin{ruledtabular}
\begin{tabular}{|ll|}
\multicolumn{2}{|c|}{\textbf{Fastball Group ($G_0$)}}               \\ \hline
\textbf{FF} – Fastball (4-seam) & \textbf{SI} – Sinker (2-seam)     \\
\textbf{FC} – Cutter            &                                   \\ \hline \hline

\multicolumn{2}{|c|}{\textbf{Off-speed Group ($G_1$)}}               \\ \hline
\textbf{CH} – Changeup          & \textbf{FS} – Split-finger        \\  
\textbf{FO} – Forkball          & \textbf{SC} – Screwball           \\ \hline \hline

\multicolumn{2}{|c|}{\textbf{Breaking Group – Curveballs ($G_2$)}} \\ \hline
\textbf{CU} – Curveball         & \textbf{KC} – Knuckle Curve      \\
\textbf{CS} – Slow Curve        &                                  \\ \hline \hline

\multicolumn{2}{|c|}{\textbf{Breaking Group – Sliders ($G_3$)}}    \\ \hline
\textbf{SL} – Slider            & \textbf{ST} – Sweeper            \\ 
\textbf{SV} – Slurve            &                                  \\ \hline \hline

\multicolumn{2}{|c|}{\textbf{Knuckleball ($G_4$)}}                 \\ \hline
\textbf{KN} – Knuckleball       &                                  \\ \hline \hline

\multicolumn{2}{|c|}{\textbf{Other Pitches ($G_5$)}}               \\ \hline
\textbf{EP} – Eephus            & \textbf{FA} – Other              \\
\textbf{IN} – Intentional Ball  & \textbf{PO} – Pitchout           \\
\end{tabular}
\end{ruledtabular}
\end{table}

\begin{figure*}[t]
\includegraphics[width=0.95\linewidth]{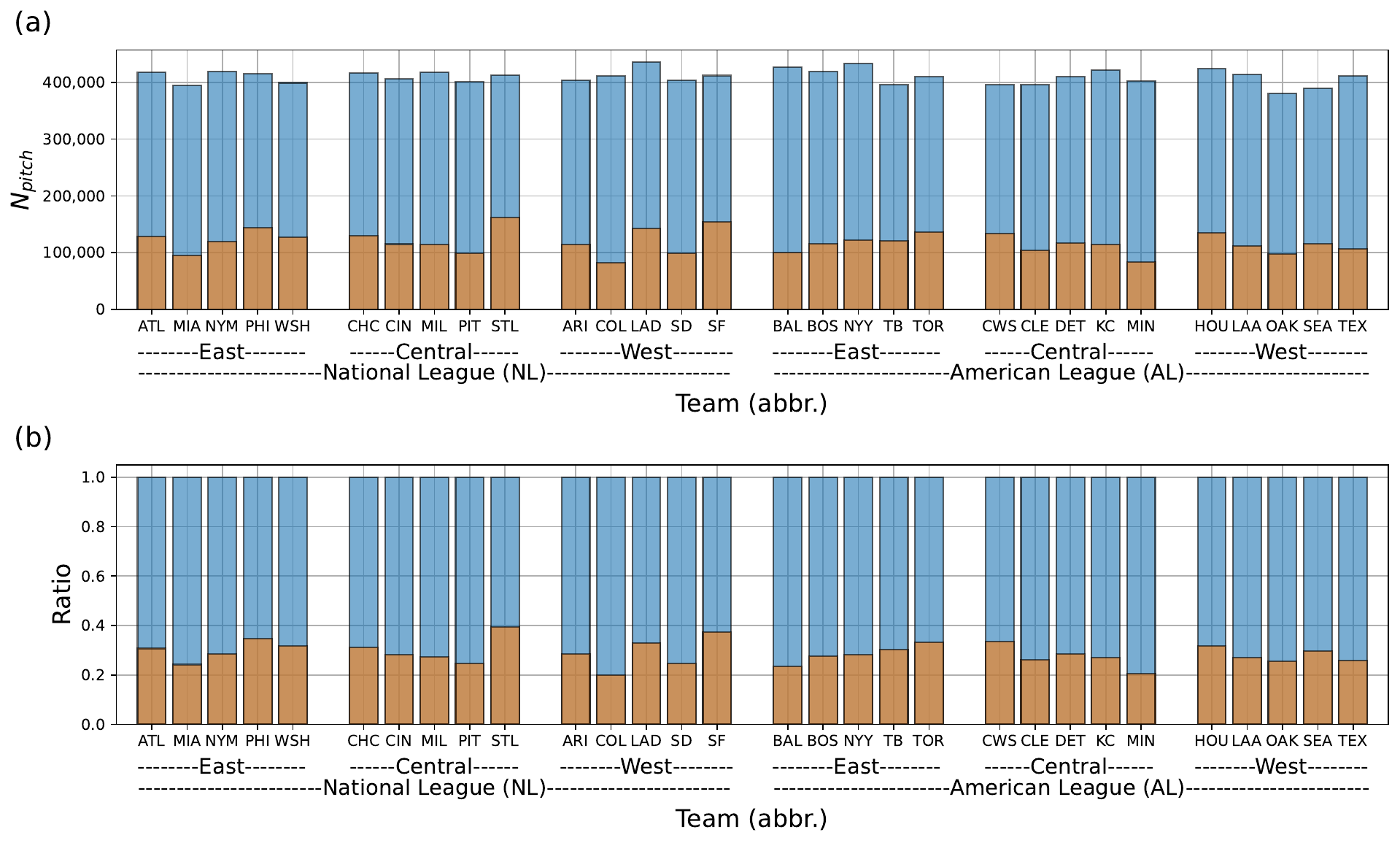}
\caption{\textbf{The number of pitches across individual teams.} The MLB consists of two leagues (the National League and the American League), each divided into three divisions (East, Central, and West) with five teams per division, for a total of 30 teams (see Table~\ref{table:mlb_team}). The proportions of pitches thrown by pitchers who recorded at least 162 innings in a given season are shown: (a) the total number of pitches $N_{\rm pitch}$ and (b) the corresponding proportion. In both panels, the combined height of the blue and orange bars indicates the total number of pitches per team, while the orange bars represent the number of pitches thrown by qualified pitchers. From 2008 to 2025, teams averaged approximately 400,000 pitches per season, with about 30\% contributed by pitchers meeting the qualification threshold.}
\label{fig:n_pitch_team}
\end{figure*}

\begin{figure*}[t]
\includegraphics[width=\linewidth]{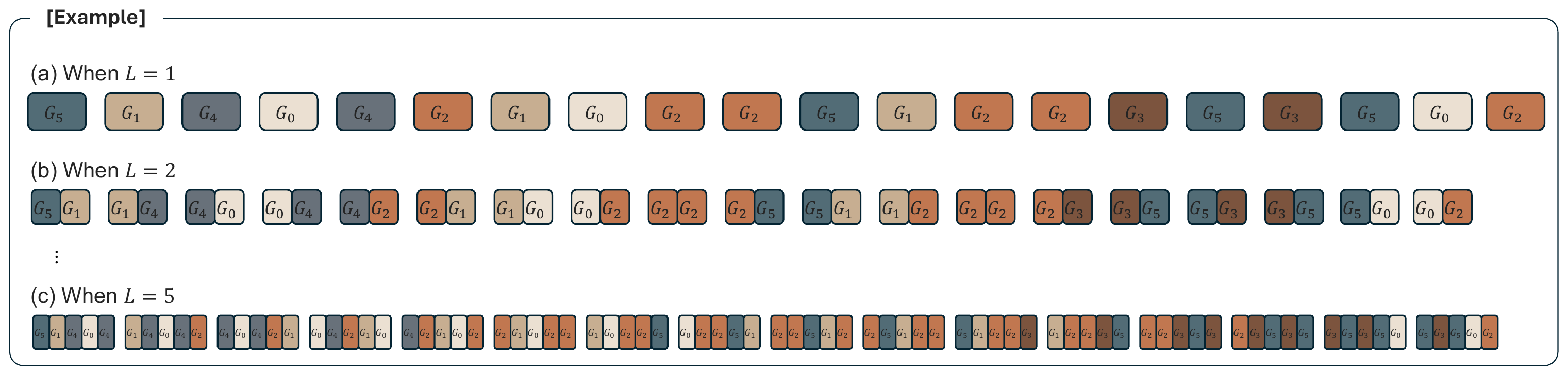}
\caption{\textbf{Schematic illustration of pitch-pattern motif construction.} The pitch-pattern motif is constructed as follows: (i) the 20 Statcast pitch sequence are remapped into the six groups listed in Table~\ref{table:pitch_type} to simplify the analysis; (ii) each pitch sequence is swept with a sliding window of length $L$ (panels (a--c) illustrate examples for $L=1,~2$, and 5) to enumerate all contiguous $L$-pitch strings; and (iii) the resulting unique strings are treated as pitch-pattern motifs that form the basis of the information-theoretic analysis of pitch usage.}
\label{fig:motif_definition}
\end{figure*}

To ensure consistency in long-term sequence analysis, we filter the datasets according to two criteria: (i) retaining only pitchers who recorded at least 162 innings pitched (the MLB standard for ``qualified'' status) and (ii) excluding the year 2020. Although this criterion introduces a sampling bias toward starting pitchers, the 162-inning threshold is widely used in professional-league analyses~\cite{mlb_dot_com} because it provides sufficiently long sequences for robust statistical analysis. From 2008 to 2025, the total seasonal number of pitches, $N_{\rm pitch}$, remains approximately 700,000 per year, with the notable exception of the pandemic-shortened 2020 season (see Fig.~\ref{fig:n_pitch_year}). The fraction of pitches thrown by qualified pitchers is highlighted by orange bars, accounting for roughly 15.8\% (2021) to 41.2\% (2011) of the league-wide total. After filtering, the final dataset comprises 3,542,338 pitches thrown by 393 qualified pitchers (see Table~\ref{table:overview}).

MLB consists of the National League (NL) and the American League (AL), each subdivided into East, Central, and West divisions with five clubs per division, for a total of 30 teams (see Table~\ref{table:mlb_team}). At the team level, individual clubs record approximately 400,000 pitches per year, of which about 30\% are thrown by pitchers who meet the qualified-inning criterion (see Fig.~\ref{fig:n_pitch_team}).

\begin{figure*}[t]
\includegraphics[width=\linewidth]{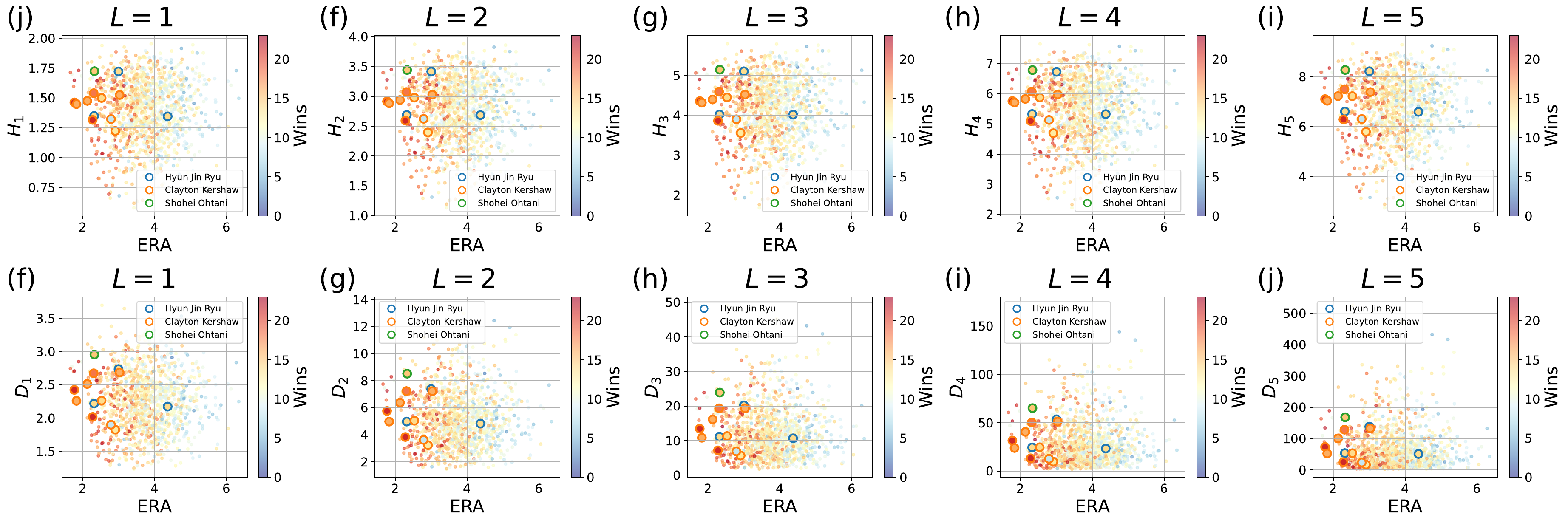}
\caption{\textbf{Pitch-pattern motif information across motif length.} (a-e) the Shannon entropy, $H_L$, of pattern usage for motif lengths $L=1\dots5$, (f–j) the inverse Simpson diversity index, $D_L$, for the same lengths. Panel pairs share a length: (a, f) $L=1$ ($G_i$ frequencies), (b, g) $L=2$, (c, h) $L=3$, (d, i) $L=4$, and (e, j) $L=5$. Each point indicates one of the 393 pitchers who meet the qualified threshold (at least 162 innings pitched), a requirement that keeps the sequences long enough for dependable diversity estimates. The horizontal axis indicates the ERA, and the wins set the color scale; the lower ERA values correspond to the higher wins, yet neither $H_L$ nor $D_L$ shows a clear link to those statistics.}
\label{fig:era_pattern_all}
\end{figure*}

\section{Methods}
\label{sec:method}
Release speed directly shapes hitter timing and provides a consistent lens, in contrast to subjective pitch names. When a repertoire combining fastballs with breaking or off-speed pitches is grouped by speed ranges, pitching strategies can be examined within an objective pattern space, avoiding reliance on pitcher-specific and subjective pitch-type labels. To simplify pitch-pattern motif analysis, we adopt the pitch-type classification provided by Statcast~\cite{Statcast}. In the Statcast interface, 18 pitch types are grouped into six categories based on internal classification criteria: fastball group, off-speed group, curveball group, slider group, knuckleball, and other pitches. We analyze pitch sequence with these six categories as listed in Table~\ref{table:pitch_type}. This mapping reduces the pitch-pattern motif space from $18^L$ to $6^L$ while preserving the dominant strategic structures.

For each pitcher, we arrange the season’s pitch sequence chronologically and extract temporal pitch-pattern motifs using a sliding window of length $L\in\{1,2,3,4,5\}$ applied to the coarse-grained pitch groups listed in Table~\ref{table:pitch_type}. Let $p_{L,i}$ denote the empirical probability of observing pitch-pattern motif $i$ of length $L$, computed as its relative frequency within the sequence. For example, when $L=1$, the motifs correspond directly to the original temporal sequence of coarse-grained pitch groups. In Fig.~\ref{fig:motif_definition}(a), the probability of pitch-pattern motif $i=0$ is $p_{L=1,i=0}=3/20$. When $L=2$, temporally adjacent pairs of pitch groups are treated as single motifs, yielding a total of 19 motifs. The probability of the $G_3$-$G_5$ pitch-pattern motif is then evaluated as $2/19$. For notational convenience, we encode each pitch-pattern motif in base-6 by mapping $G_0, \ldots, G_5$ to digits $0, \ldots, 5$, respectively. In this notation, for example, the first $L=5$ pitch-pattern motif in Fig.~\ref{fig:motif_definition}(c), $(G_5,G_1,G_4,G_0,G_4)$, is written as $51404_{(6)}$. We use this base-6 notation consistently in the following tables and figures.

To quantify the temporal diversity of pitch patterns, we compute the Shannon entropy ($H_L$) and the inverse Simpson index ($D_L$) from the pitch-pattern motif distribution $\{p_{L,i}\}$. The Shannon entropy primarily captures the informational richness of the distribution and is defined as
\begin{equation}
H_L=-\sum_{i} p_{L,i} \log{\left(p_{L,i}\right)}~.
\end{equation}
The inverse Simpson index $D_L$ characterizes the concentration of probability mass among a limited number of dominant motifs, yielding the effective number of dominant pitch-pattern motifs, and is given by
\begin{equation}
D_L=\left(\sum_{i} \left(p_{L,i}\right)^2\right)^{-1}~.
\end{equation}
Together, these measures distinguish between broadly diversified pitch-pattern motif repertoires and sequences dominated by a small set of recurrent patterns.

Having characterized the temporal diversity of pitch-pattern motifs, we next examine how this diversity relates to commonly used measures of pitching outcomes. As a baseline comparison, we consider earned run average (ERA) and wins, which summarize run prevention and game-level outcomes at the pitcher level. ERA quantifies the number of earned runs allowed per nine innings, with lower values indicating more effective run prevention. This comparison allows us to assess whether pitch-pattern motif diversity exhibits any direct correspondence with conventional performance metrics. 

In parallel, we analyze the rank--frequency distribution of pitch-pattern motifs to probe the internal structure of pitch-pattern motif usage itself. Specifically, the rank--frequency curve provides a reference for evaluating whether observed pitch-pattern motif sequences resemble random combinations or display systematic, non-random organization~\cite{choi2005zipf,kiet2007korean}. Together, these complementary analyses distinguish between outcome-level associations and intrinsic structural properties of pitch-pattern motifs.

\begin{table}[t]
\caption{\textbf{Top 5 pitch-pattern motif of multiple lengths by outcome.}
For each sequence length, we list the five most frequent pitch-pattern motifs (base-6 coding) together with their proportions for all plate appearances (ALL), hits
(HIT), and outs (OUT). We report the top‑5 motifs for ALL, HIT, and OUT to compare outcomes and assess which patterns change most when conditioning on HIT or OUT plate appearances relative to the overall distribution. The hit and out columns show almost identical ranks and
weights, indicating that even short-to-medium pitch sequences fail to separate
successful and unsuccessful results. All statistics use the qualified dataset containing 3,542,338 pitches.}
\label{table:top_patterns_rank_all}
\begin{ruledtabular}
\begin{tabular}{c c | ccc}$L$ & Rank & ALL & HIT & OUT \\
\hline\hline
\multirow{5}{*}{2}
 & Top 1 & $    00_{(6)}$ 0.3885 & $    00_{(6)}$ 0.4040 & $    00_{(6)}$ 0.3616 \\
 & Top 2 & $    30_{(6)}$ 0.0774 & $    10_{(6)}$ 0.0819 & $    03_{(6)}$ 0.0842 \\
 & Top 3 & $    03_{(6)}$ 0.0773 & $    30_{(6)}$ 0.0816 & $    01_{(6)}$ 0.0814 \\
 & Top 4 & $    10_{(6)}$ 0.0767 & $    01_{(6)}$ 0.0789 & $    30_{(6)}$ 0.0767 \\
 & Top 5 & $    01_{(6)}$ 0.0742 & $    03_{(6)}$ 0.0700 & $    10_{(6)}$ 0.0722 \\
\hline
\multirow{5}{*}{3}
 & Top 1 & $   000_{(6)}$ 0.2575 & $   000_{(6)}$ 0.2699 & $   000_{(6)}$ 0.2424 \\
 & Top 2 & $   100_{(6)}$ 0.0463 & $   100_{(6)}$ 0.0480 & $   003_{(6)}$ 0.0483 \\
 & Top 3 & $   003_{(6)}$ 0.0448 & $   010_{(6)}$ 0.0471 & $   001_{(6)}$ 0.0481 \\
 & Top 4 & $   300_{(6)}$ 0.0447 & $   001_{(6)}$ 0.0469 & $   030_{(6)}$ 0.0431 \\
 & Top 5 & $   001_{(6)}$ 0.0446 & $   030_{(6)}$ 0.0457 & $   100_{(6)}$ 0.0413 \\
\hline
\multirow{5}{*}{4}
 & Top 1 & $  0000_{(6)}$ 0.1771 & $  0000_{(6)}$ 0.1871 & $  0000_{(6)}$ 0.1694 \\
 & Top 2 & $  1000_{(6)}$ 0.0288 & $  1000_{(6)}$ 0.0301 & $  0001_{(6)}$ 0.0294 \\
 & Top 3 & $  0001_{(6)}$ 0.0276 & $  0010_{(6)}$ 0.0294 & $  0003_{(6)}$ 0.0287 \\
 & Top 4 & $  0100_{(6)}$ 0.0273 & $  0001_{(6)}$ 0.0289 & $  0030_{(6)}$ 0.0263 \\
 & Top 5 & $  0010_{(6)}$ 0.0272 & $  0100_{(6)}$ 0.0283 & $  1000_{(6)}$ 0.0261 \\
\hline
\multirow{5}{*}{5}
 & Top 1 & $ 00000_{(6)}$ 0.1259 & $ 00000_{(6)}$ 0.1338 & $ 00000_{(6)}$ 0.1214 \\
 & Top 2 & $ 10000_{(6)}$ 0.0184 & $ 10000_{(6)}$ 0.0190 & $ 00001_{(6)}$ 0.0188 \\
 & Top 3 & $ 00001_{(6)}$ 0.0177 & $ 00010_{(6)}$ 0.0190 & $ 00003_{(6)}$ 0.0178 \\
 & Top 4 & $ 01000_{(6)}$ 0.0175 & $ 01000_{(6)}$ 0.0186 & $ 10000_{(6)}$ 0.0175 \\
 & Top 5 & $ 00100_{(6)}$ 0.0175 & $ 00001_{(6)}$ 0.0184 & $ 00020_{(6)}$ 0.0167 \\
\hline
\multirow{5}{*}{6}
 & Top 1 & $000000_{(6)}$ 0.0919 & $000000_{(6)}$ 0.0984 & $000000_{(6)}$ 0.0892 \\
 & Top 2 & $100000_{(6)}$ 0.0122 & $100000_{(6)}$ 0.0129 & $000001_{(6)}$ 0.0124 \\
 & Top 3 & $000001_{(6)}$ 0.0118 & $000010_{(6)}$ 0.0126 & $100000_{(6)}$ 0.0116 \\
 & Top 4 & $000002_{(6)}$ 0.0116 & $000001_{(6)}$ 0.0124 & $000003_{(6)}$ 0.0114 \\
 & Top 5 & $010000_{(6)}$ 0.0115 & $001000_{(6)}$ 0.0123 & $000020_{(6)}$ 0.0114 \\
\end{tabular}
\end{ruledtabular}
\end{table}

\begin{figure}[t]
\includegraphics[width=\linewidth]{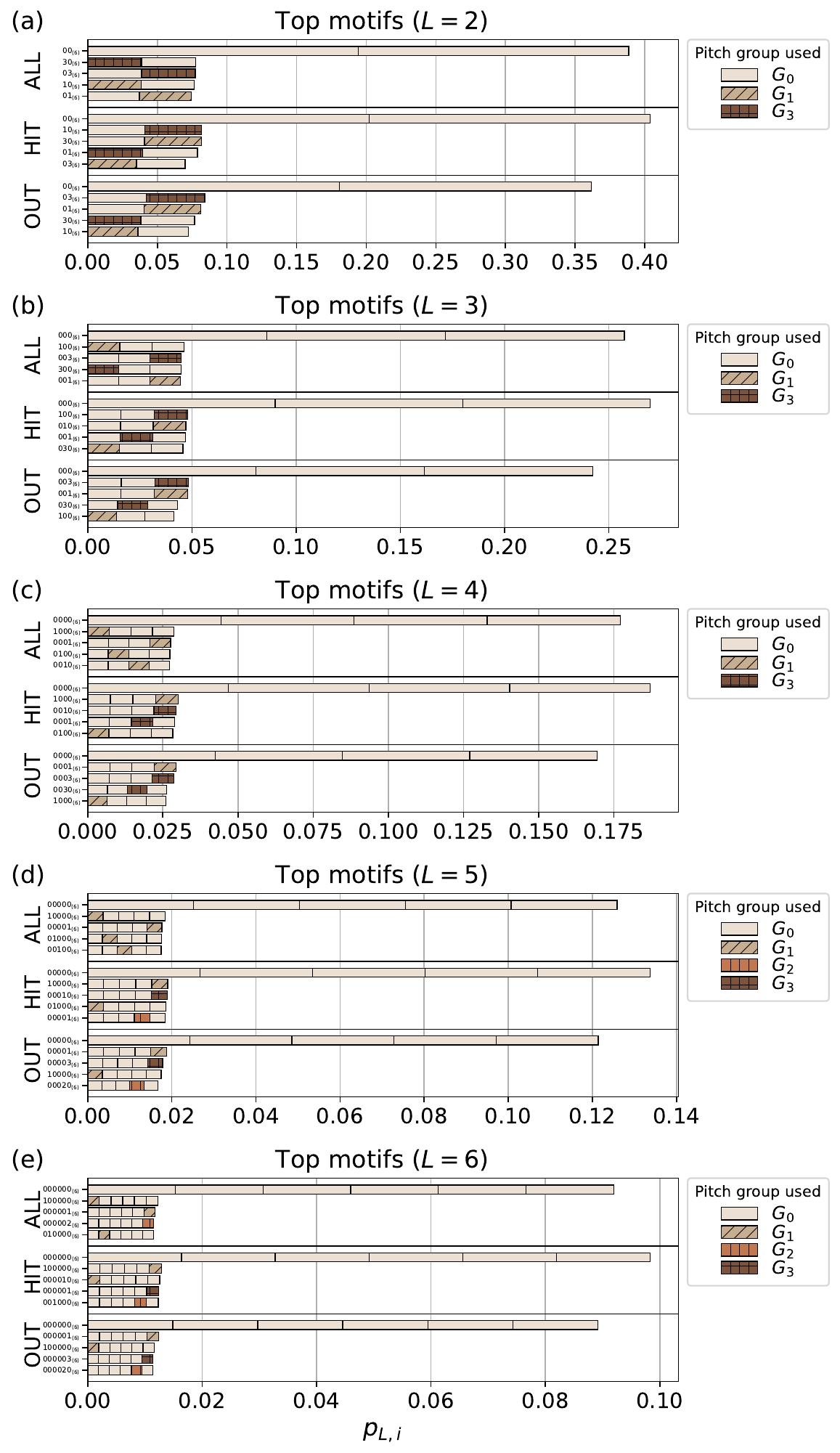}
\caption{\textbf{Visualization of the five most frequent pitch-pattern motifs shown in Table~\ref{table:top_patterns_rank_all}}. The horizontal length of each bar corresponds to the probability $p_{L,i}$, and each bar is divided into $L$ blocks. The sequence of blocks from the leftmost position represents a number encoded in base-6, with each digit indicated by color and pattern. For example, the bar ranked fifth for OUT in the case of $L=2$ [panel (a)] corresponds to $02_{(6)}$.}
\label{fig:seq}
\end{figure}

\section{Results}
\label{sec:results}
\subsection{Pitcher-level analysis (macroscopic level)}
\label{subsec:macro}
To explore whether pitch-sequence structure is related to pitcher performance, we evaluate the motif-length--dependent Shannon entropy ($H_L$) and inverse Simpson diversity ($D_L$), and compare these values with representative performance indicators such as ERA and wins.

Figure~\ref{fig:era_pattern_all} shows comparisons of $H_L$ [panels (a--e)] and $D_L$ [panels (f--j)] against ERA, with seasonal wins, for the 393 qualified pitchers. The horizontal axis encodes ERA, while the color scale represents total wins; panel pairs [(a, f), (b, g), $\cdots$, (e, j)] correspond to identical pitch-pattern motif lengths $L=1,\cdots,5$. As expected, ERA and wins exhibit a negative correlation: as ERA decreases (i.e., as pitchers allow fewer earned runs), seasonal wins naturally tend to increase. In contrast, both $H_L$ and $D_L$ remain clustered within relatively narrow ERA ranges, without an appreciable change in dispersion as $L$ increases. The diversities of pitchers with contrasting styles and profiles (e.g., Hyun Jin Ryu, Clayton Kershaw, or Shohei Ohtani) are generally high, but tend to be scattered rather than forming well-defined clusters. This suggests that relating pitch-sequence diversity to performance may require performance indicators independent of ERA or wins, and vice versa. Overall, pitch-sequence diversity shows no clear association with pitcher performance at the level of these outcome metrics.

\subsection{Pitch-pattern motif-level analysis (microscopic level)}
We now implement pitch-pattern motif-level analysis on the pitches in the qualified dataset, rather than restricting attention to pitcher-level aggregated motif statistics. This analysis covers $N_{\rm pitch}=3,542,338$ pitches recorded by Statcast between 2008 and 2025. From this microscale perspective, we examine whether local pitch sequences are associated with hit (success) or out (failure) events, which may not be captured by the macroscopic diversity measures $H_L$ and $D_L$.

Table~\ref{table:top_patterns_rank_all} lists the five most frequent motifs encoded in base-6, reflecting the six coarse-grained pitch groups, for lengths $L=2,\dots,6$ (pitch-type encoding with frequencies or probabilities $p_{L,i}$ listed in Table~\ref{table:pitch_type}). For instance, $012_{(6)}$ indicates the sequence $G_0, G_1$, and $G_2$. The pitch-pattern motifs encoded in base-6 are categorized into hits (HIT), outs (OUT), and all plate appearances (ALL), where ALL includes both HIT and OUT cases. The table is visualized in Fig.~\ref{fig:seq}. Very similar sequences frequently appear in both HIT and OUT categories, indicating that no decisive or privileged pitch sequence systematically determines outcomes as HIT or OUT.
    
Although a keystone pitch-pattern motif sequence is absent for determining HIT or OUT, one can note that the three groups $G_0, G_1,$ and $G_3$ appear most frequently. These groups correspond to the fastball group ($G_0$), the off-speed group ($G_1$), and the breaking group--sliders ($G_3$). In particular, pitchers overwhelmingly exploit pitch types belonging to the fastball group $G_0$. This heavy use of $G_0$ highlights the extent to which MLB pitchers rely on a limited set of recurrent patterns (see Appendix~\ref{sec-app:pattern} for detailed distributions).

Fastballs $G_0$ constitute the indispensable baseline of pitch sequencing, establishing the reference speed and command against which all subsequent variation is interpreted. Without this baseline, a pitch sequence lacks both physical coherence and strategic meaning. Off-speed pitches $G_1$ derive their effectiveness precisely from this reference, as their timing-disrupting role is inherently relative to fastballs rather than effective in isolation. Sliders $G_3$ further complement this structure by providing a low-cost mode of variation: they introduce pronounced lateral movement while maintaining near-fastball velocity and mechanical continuity. Together, these three pitch groups form a minimal yet sufficient set, naturally dominating usage as the core components required to sustain physically and strategically viable pitch sequences across diverse game situations.

\begin{figure}[t]
\includegraphics[width=\linewidth]{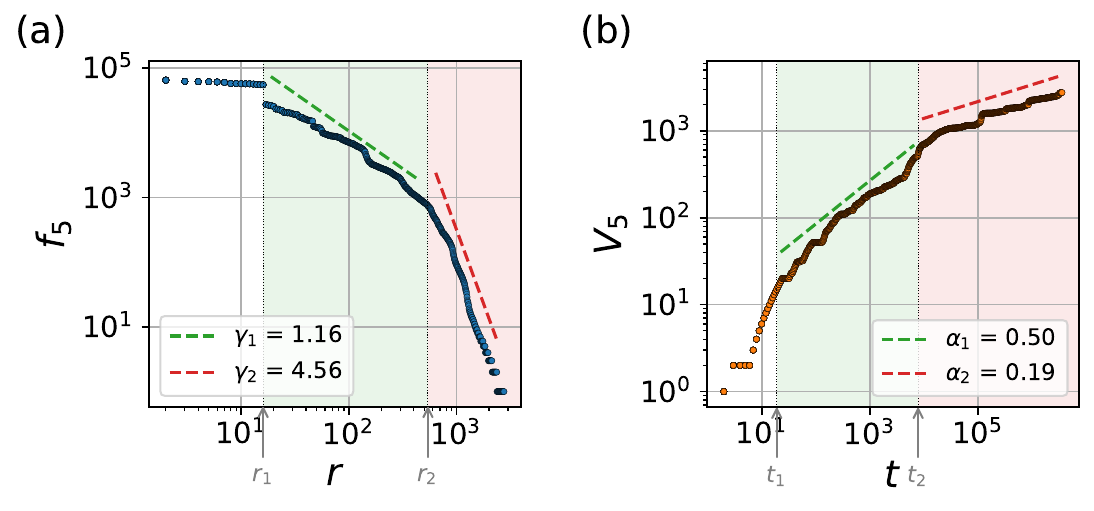}
\caption{\textbf{Zipf and Heaps plots for length-5 pitch-pattern motifs.} (a) Rank--frequency plot for length-5 pitch-pattern motifs. The intermediate range $r_1 \lesssim r \lesssim r_2$ (green region) shows $f_{5}(r)\sim r^{-1.2}$ (Zipf-like behavior), while the exponent changes from $\gamma_{1}\simeq 1.2$ to $\gamma_{2}\simeq 4.6$ beyond the cutoff $r_2$ (red region). The crossover points $r_1$ and $r_2$ are identified using the Kolmogorov--Smirnov statistic. (b) Growth of the vocabulary size $V_{5}(t)$ (the number of distinct pitch-pattern motifs) as a function of the number $t$ of tokens. The scaling boundaries $t_1$ and $t_2$ are determined by $V_{5}(t_1)\simeq r_1$ and $V_{5}(t_2)\simeq r_2$ [see the main text]. In $V_{5}(t)\sim t^{\alpha}$, $\alpha_{1}\simeq 0.50$ for $t_1 \lesssim t \lesssim t_2$ (Heaps-like behavior), while $\alpha_{2}\simeq 0.19\approx 1/\gamma_{2}$ for $t \gtrsim t_2$. All exponents are estimated by least-squares fitting in log-log space.}
\label{fig:zipf_heaps_pattern5}
\end{figure}

Although no $L$-motif sequence uniquely determines HIT or OUT outcomes, the repeated use of specific pitch patterns suggests the presence of stable usage rules. To probe such structure, we analyze the rank--frequency distribution of $L$-motifs via Zipf’s law and their growth behavior via Heaps’ law, focusing on representative 5-motifs ($L=5$).

In language studies, Zipf’s and Heaps’ laws are commonly used to characterize the structural organization of symbolic sequences, independent of semantic meaning or communicative success. Zipf’s law captures how unevenly tokens are distributed across different types, while Heaps’ law quantifies how the number of distinct types grows as the sequence length increases. Applied to pitch sequencing, these tools test whether $L$-motif usage follows systematic, non-random constraints, thereby probing the underlying organization of pitch sequences beyond their ability to distinguish hits from outs.

To explore Zipf’s law, we employ the rank--frequency plot. Since the probability $p_{L,i}$ is equivalent to the relative frequency, we define the pitch-pattern motif frequency as $f_{L,i}\equiv p_{L,i} N_L$, where $N_L$ denotes the total number of temporal motifs of length $L$ (see Fig.~\ref{fig:motif_definition}). We then order motifs in descending frequency and define
\begin{equation}
  \begin{aligned}
  f_L(r) \equiv {} & \text{the frequency of the pitch-pattern motif} \\
                   & \text{with rank } r .
  \end{aligned}
  \label{eq:freq}
\end{equation}
For Heaps’ law, we define a token as a single occurrence of an $L$-length pitch-pattern motif extracted from the pitch sequence. The vocabulary size (borrowing terminology from language analysis) is given by
\begin{equation}
    V_L(t)\equiv \text{the number of distinct tokens for } t \text{ tokens}.
\end{equation}
For example, consider the sequence $(G_0, G_1, G_3, G_0, G_1)$. When $L=2$, the extracted tokens are $(G_0, G_1)$, $(G_1, G_3)$, $(G_3, G_0)$, and $(G_0, G_1)$, yielding $t=4$ and $V_L(t)=3$. In language studies, the scaling relations
\begin{align}
    f_L(r)&\sim r^{-\gamma}, \label{eq:fr}\\
    V_L(t)&\sim t^{\alpha} \label{eq:vt}
\end{align}
are commonly examined. The exponents $\gamma$ and $\alpha$ are related through
$t\sim\int_{1}^{V} f(r)\,dr \sim \int_{1}^{V} r^{-\gamma}\,dr$, yielding $\gamma=1/\alpha$ for $\alpha>1$. Empirically, $\gamma\simeq1$ and $\alpha\simeq0.5$ are typically observed, with logarithmic corrections.

The Zipf and Heaps plots for the length-5 motifs are shown in Fig.~\ref{fig:zipf_heaps_pattern5}, where power-law-like patterns are observed. In Fig.~\ref{fig:zipf_heaps_pattern5}(a), two kinks are observed at $r=r_1$ and $r=r_2$, identified via the Kolmogorov-Smirnov statistic. These rank scales correspond to characteristic ranges $t_1$ and $t_2$ in the Heaps plot [Fig.~\ref{fig:zipf_heaps_pattern5}(b)]. This correspondence can be understood from the relation between rank and vocabulary growth. Let $p_r\equiv f(r)/t_{\rm total}$ denote the probability of the motif ranked $r$, where $t_{\rm total}$ is the total number of tokens. The expected vocabulary size is $\mathbb{E}[V(t)] = \sum_r [1-(1-p_r)^t] \approx \sum_r (1-e^{-t p_r})$ for small $p_r$. The exponential term is effectively 1 for $t p_r \gg 1$ and 0 for $t p_r \ll 1$, with the crossover defined by $t p_r^{*}\sim1$, leading to $\mathbb{E}[V(t)] \approx r^*$. Consequently, the crossover points satisfy $V_{5}(t_1)\approx r_1$ and $V_{5}(t_2)\approx r_2$.

Consistent with this relation, Fig.~\ref{fig:zipf_heaps_pattern5} shows that $f_{5}(r)$ and $V_{5}(t)$ follow Zipf’s and Heaps’ laws over intermediate ranges, with $\gamma_{1}\simeq1$ for $r_1 \lesssim r \lesssim r_2$ and $\alpha_{1}\simeq0.5$ for $t_1 \lesssim t \lesssim t_2$, similar to those observed in language systems. The deviation for $r\gtrsim r_2$ (corresponding to $t\gtrsim t_2$) arises from finite-size effects, while the relation $\gamma_{2}=1/\alpha_{2}$ remains approximately valid ($\gamma_{2}\approx 5$ and $\alpha_{2}\approx 0.2$). The intermediate scaling range, over which Zipf-like and Heaps-like behaviors are observed, is expected to broaden with increasing system size.

These results indicate that pitch-pattern motif usage is not random. Under random sampling, the Zipf plot would approach a nearly flat profile with diminishing frequency contrast across ranks, and the Heaps curve would rapidly saturate toward a fixed vocabulary size. In contrast, the observed scaling behavior reflects persistent heterogeneity and sustained vocabulary growth, consistent with language-like organization in pitch sequencing.

Overall, neither $H_L$ nor $D_L$ exhibits a strong correlation with conventional performance measures such as wins or ERA across the MLB sample. Pitchers with comparable entropy or diversity values nevertheless span a wide range of outcomes, indicating that sequence-level diversity alone does not uniquely characterize pitching success. Rather, these results suggest that performance is likely mediated by additional conditional factors---such as count leverage, velocity differentials, or pitch location---that are not captured by motif-based summaries. At the same time, the presence of stable motif usage and language-like scaling indicates that pitch sequencing is governed by non-random structural constraints. Together, these findings delineate both the scope and the limits of sequence-based metrics, naturally motivating a discussion of how contextual information may complement motif-level analyses.

\section{Discussion}
\label{sec:discusssion}
We have analyzed pitch-pattern motifs constructed from coarse-grained pitch types and examined their sequential structure. Our results show that motif-based sequence diversity at a pitcher level, as quantified by the Shannon entropy $H_L$ and the inverse Simpson index $D_L$, exhibits less decisive association with conventional performance measures such as ERA or wins. Likewise, even the most highly ranked motifs do not contribute to discriminating between HIT and OUT outcomes. These findings indicate inherent limitations in using single pitch-pattern motifs, or motif-based diversity measures alone, as independent variables for predicting seasonal pitching performance.

Nonetheless, pitch usage exhibits internally organized structure that emerges at the macroscopic level, despite the absence of outcome-determining sequences~\cite{park2021motif}. This interpretation is supported by the observation that both Zipf-like and Heaps-like behaviors hold for pitch-motif usage, indicating systematic, non-random organization. While individual pitchers operate with distinct styles and preferences, the aggregated motif statistics reveal a robust, scale-invariant usage pattern shared across the league. In this sense, pitch sequencing displays grammar-like regularities analogous to those observed in language, not as a semantic system, but as a structured rule-governed process~\cite{yoon2005distributions,park2017network,park2018korean,kim2018quantitative}.

Although baseball is inherently a sport driven by pitcher--batter interactions, this study does not explicitly consider such interactions in the pitch-pattern analysis, primarily due to limitations in data availability. Accordingly, the present conclusions primarily reflect pitcher-level, interaction-averaged pitching patterns. To illustrate this limitation, we construct a pitcher--batter bipartite network (see Fig.~\ref{fig:bipartite_network} in Appendix~\ref{sec-app:bipartite}). The interactions between pitchers and batters are represented by edge weights, defined as the number of pitches shared. The weight distribution shown in the inset of Fig.~\ref{fig:bipartite_network} follows an approximately exponential form with a mean of $23.28$. In practice, an average pitcher--batter pair shares only about 20 pitches, which is insufficient for statistically reliable interaction-level inference. For this reason, we focus on pitcher-level information and corpus-wide motif statistics; interaction-based motif analysis would become feasible with denser pitcher--batter encounter data.

To more precisely capture how pitch-pattern motifs interact with game context, several directions for extension are required. First, conditional models exploiting count leverage, velocity differentials, and pitch location would allow the context dependence of motif-level information to be quantified. Second, a game-theoretic framework that explicitly incorporates batter-specific responses could model the strategic dynamics of pitcher--batter interactions. Third, motif-based features introduced in this study could be integrated into sequential models such as LSTM networks, enabling a direct comparison of predictive performance against established baselines and testing whether the inferred “pitch grammar” translates into measurable performance gains. These extensions are inherently sensitive to data scale and quality; consequently, long-term progress will require denser sensor recordings and broader samples of pitcher--batter encounters, including interleague play.

\section*{Acknowledgments}
\label{sec:acknowledgements}
This work was supported by a 2-Year Research Grant of Pusan National University (M.J.L.).

%

\def\tb{\textbackslash}

\onecolumngrid

\clearpage
\newpage

\appendix
\setcounter{section}{1}
\renewcommand{\thesection}{A}
\setcounter{figure}{0}
\renewcommand{\thefigure}{A.\arabic{figure}}

\section{Pitcher--batter bipartite network}
\label{sec-app:bipartite}
\begin{figure*}[h]
\includegraphics[width=\linewidth]{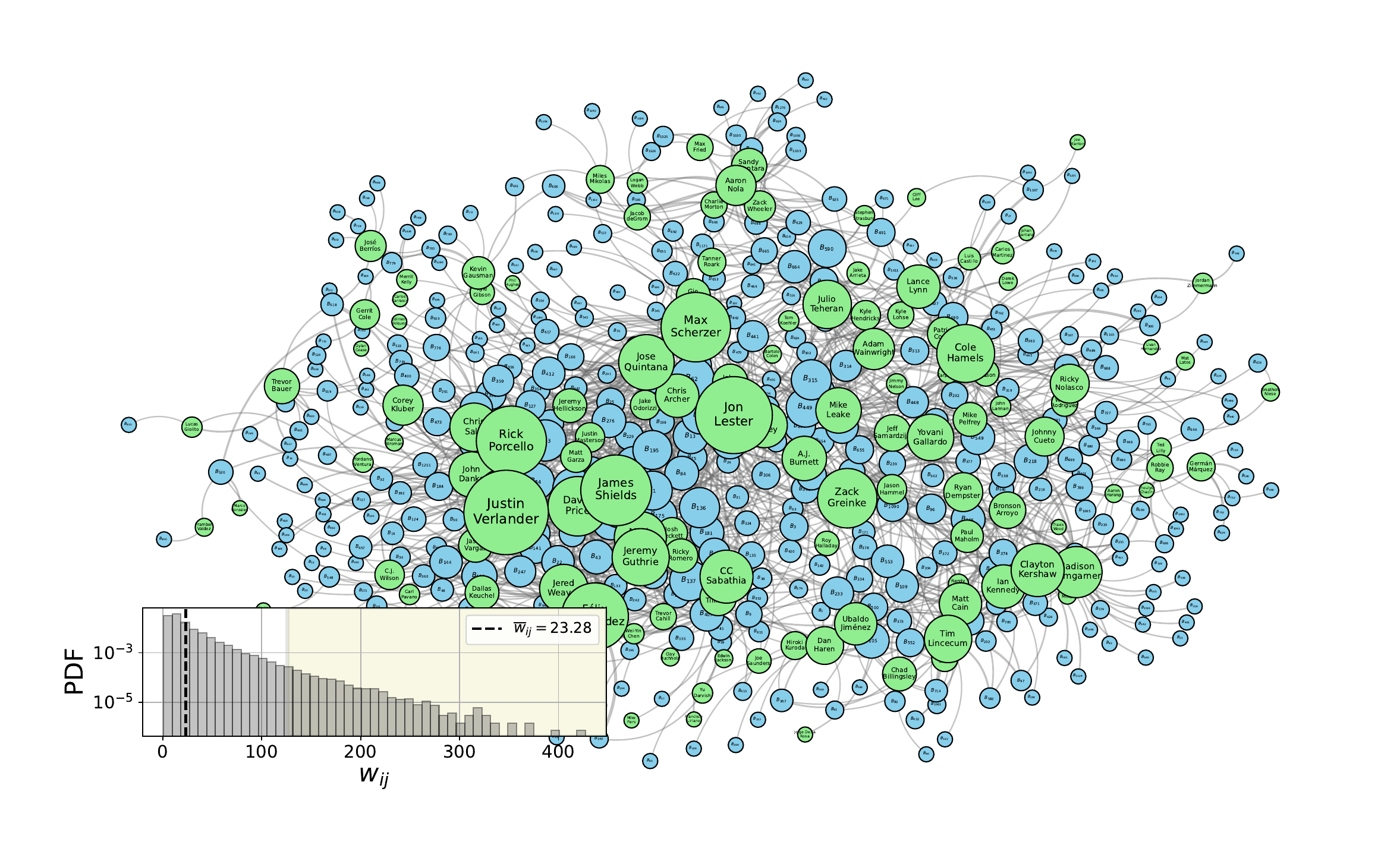}
\caption{\textbf{Pitcher–batter bipartite network.} Nodes are split into pitchers (green) and batters (sky), with batter labels shortened to $B_j$ because the analysis focuses on pitcher-centric aspects. Each edge weight, $w_{ij}$, represents the number of pitches shared by pitcher $i$ and batter $j$. Node sizes are proportional to total weights (node strengths), and edge widths scale with $w_{ij}$. For visibility, only the top 1\% of edges are shown, yielding 130 pitchers, 381 batters, and 1,544 edges out of the full-sized network. The inset shows the weight distribution and highlights the threshold used for network visualization.}
\label{fig:bipartite_network}
\end{figure*}

We construct a pitcher--batter bipartite network consisting of 393 pitchers and 3,473 batters (nodes) and 152,189 edges (Fig.~\ref{fig:bipartite_network}). Each edge weight $w_{ij}$ represents the number of pitches shared between pitcher $i$ and batter $j$. As shown in the inset of Fig.~\ref{fig:bipartite_network}, the resulting weight distribution is approximately exponential, with a mean value of $\bar{w}\simeq23.28$. In practice, an average pitcher--batter pair shares only about 20 pitches, which is insufficient for statistically reliable estimation of outcome dependence or adaptive sequencing strategies. For this reason, we focus on pitcher-level information and corpus-wide motif statistics in this study. Nevertheless, interaction-based motif analysis would become feasible, provided that more extensive pitcher--batter encounter data are accumulated.

\renewcommand{\thesection}{B}
\setcounter{figure}{0}
\renewcommand{\thefigure}{B.\arabic{figure}}

\section{Strike-ratio across pitch-pattern motifs}
\label{sec-app:pattern}
In this section, we report motif-level pattern statistics for motif lengths $L=1,2,\cdots,5$. For each length, we examine three measurements to explore potential pitch patterns associated with game outcomes.

\begin{figure*}[h]
\includegraphics[width=0.86\linewidth]{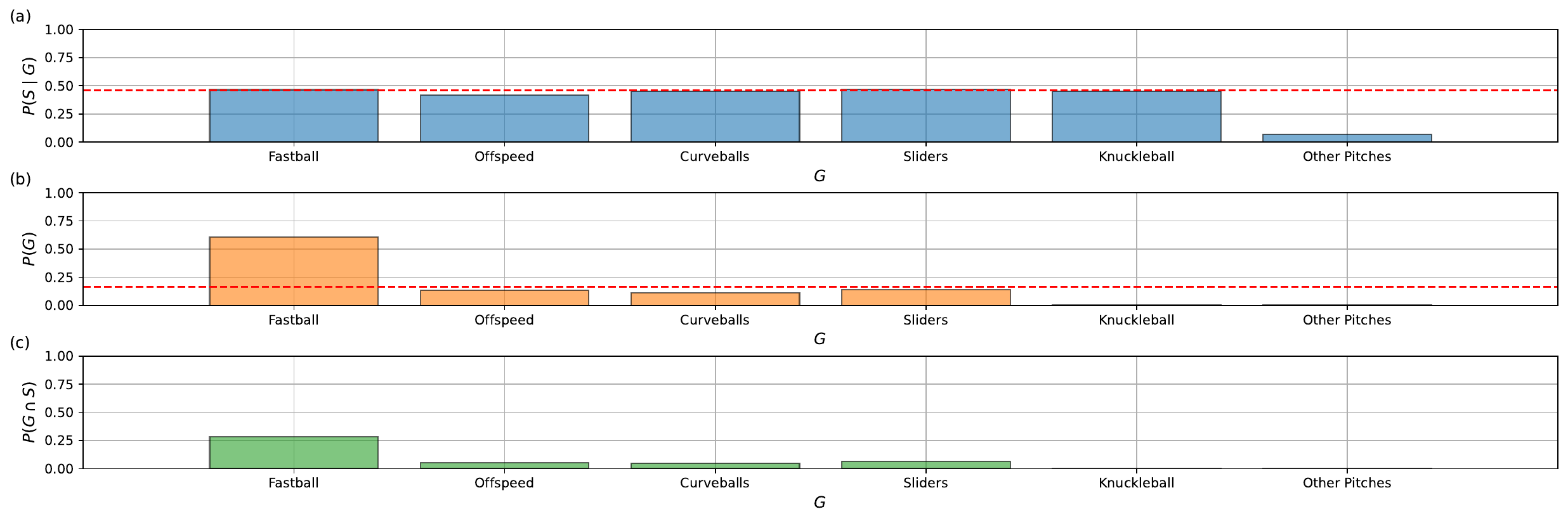}
\caption{\textbf{Pattern-level metrics of pitch sequences with motif length 1.} (a) Strike probability $P(S|G)$ for a given pitch group $G$, with the expected value indicated by the red dashed line. (b) Probability $P(G)$ of pitch group $G$ showing the relative frequency of each pattern among all pitch sequences (red dashed line: expected value). (c) Conditional probability $P(G \cap S)$ that the pitch pattern $G$ is assessed as the strike. Each pattern corresponds to a motif of length 1, representing four consecutive pitches. All statistics use the qualified dataset containing 3,542,338 pitches.}
\label{fig:pattern_1}
\end{figure*}

Let $G$ denote the set of pitch groups ($G_i$ with $i\in\{0,1,2,3,4,5\}$ listed in Table~\ref{table:pitch_type}), $S$ represent the set of strike events (pitch outcomes), and $N(A)$ be the cardinality of a set $A$, i.e., $N(A)=|A|$.

We measure three quantities: (i) the conditional probability $P(S\mid G)$ of a strike given a pitch group $G$, which represents the probability that a pitch belonging to group $G$ is evaluated as a strike and is used to assess command consistency at the pitch-group level; (ii) the probability $P(G)$ of a pitch group $G$, which records how often each coarse-grained group occurs among all pitches and reflects repetitive sequencing tendencies; and (iii) the joint probability $P(G \cap S)$, which represents the fraction of all pitches that simultaneously belong to pitch group $G$ and are evaluated as strikes, enabling direct comparisons of group-level outcomes. These quantities are written as follows:
\begin{equation}
P(S\mid G) = \frac{N(G \cap S)}{N(G)} = \frac{P(G \cap S)}{P(G)}~,
\end{equation}
\begin{equation}
P(G)=\frac{N(G)}{N}~,
\end{equation}
\begin{equation}
P(G \cap S)=\frac{{N}(G \cap S)}{N},
\end{equation}
where $N$ is a normalization constant. 

\begin{figure*}[t]
\includegraphics[width=0.86\linewidth]{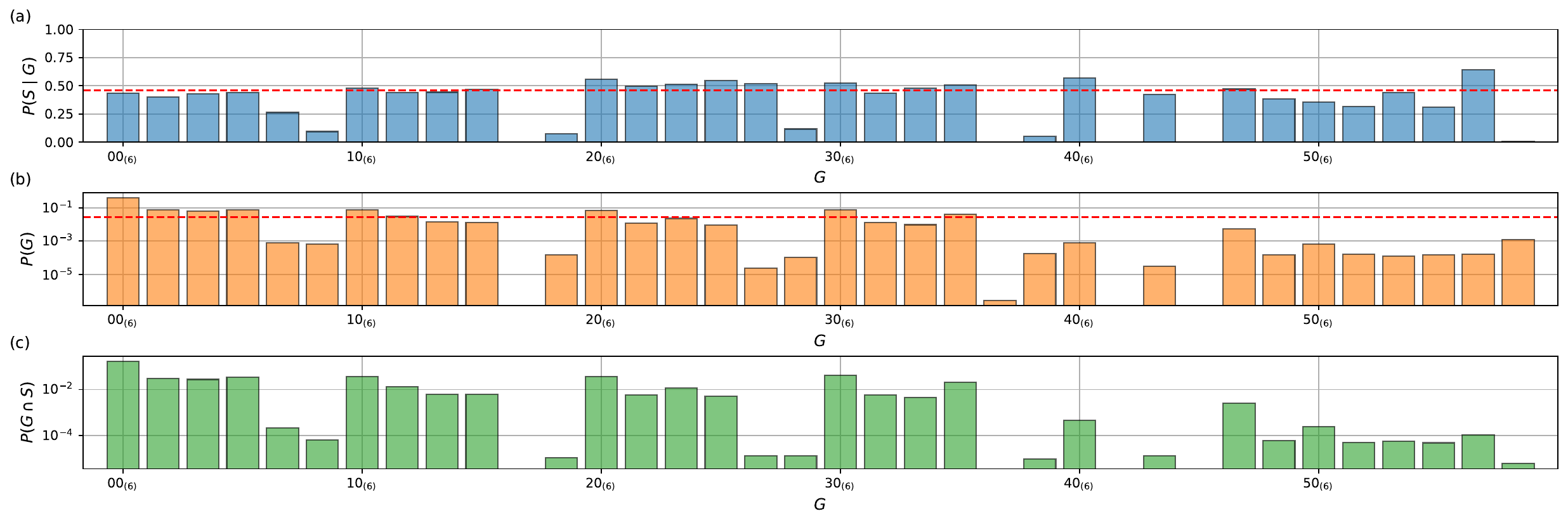}
\caption{\textbf{Pattern-level metrics of pitch sequences with pitch-pattern motif length 2.} (a) Strike probability $P(S|G)$ for a given pitch group $G$, with the expected value indicated by the red dashed line. (b) Probability $P(G)$ of pitch group $G$ showing the relative frequency of each pattern among all pitch sequences (red dashed line: expected value). (c) Conditional probability $P(G \cap S)$ that the pitch pattern $G$ is assessed as the strike. Each pattern corresponds to a motif of length 2, representing four consecutive pitches. All statistics use the qualified dataset containing 3,542,338 pitches.}
\label{fig:pattern_2}
\end{figure*}

\begin{figure*}[t]
\includegraphics[width=0.86\linewidth]{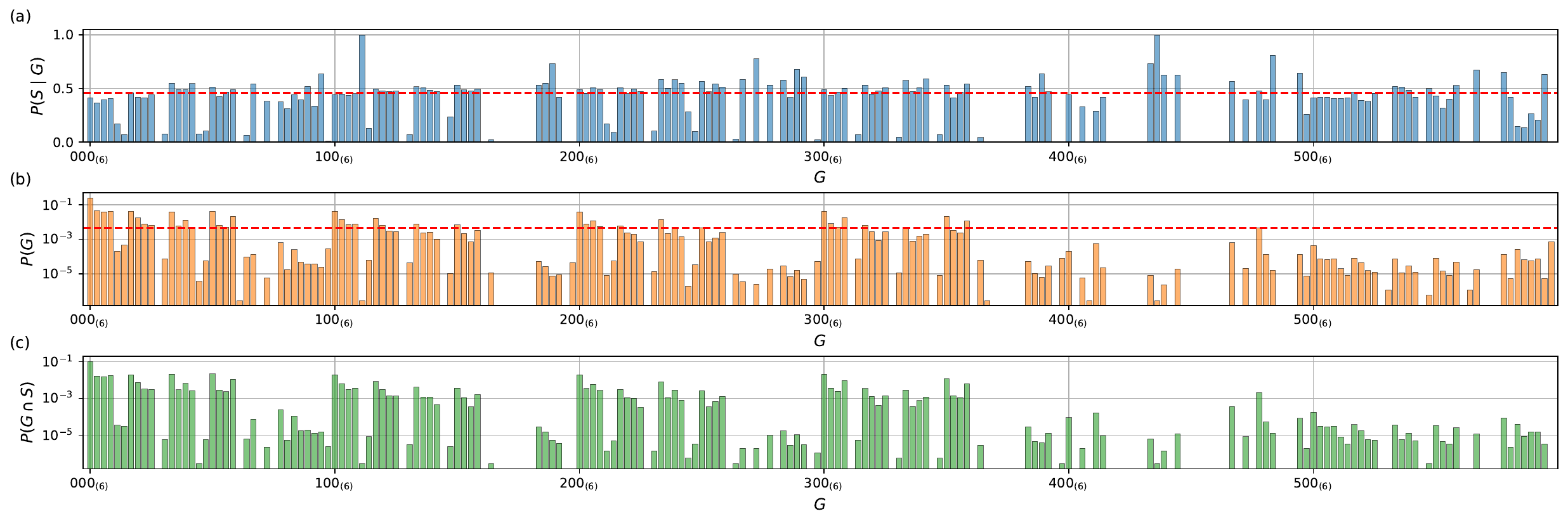}
\caption{\textbf{Pattern-level metrics of pitch sequences with pitch-pattern motif length 3.} (a) Strike probability $P(S|G)$ for a given pitch group $G$, with the expected value indicated by the red dashed line. (b) Probability $P(G)$ of pitch group $G$ showing the relative frequency of each pattern among all pitch sequences (red dashed line: expected value). (c) Conditional probability $P(G \cap S)$ that the pitch pattern $G$ is assessed as the strike. Each pattern corresponds to a motif of length 3, representing four consecutive pitches. All statistics use the qualified dataset containing 3,542,338 pitches.}
\label{fig:pattern_3}
\end{figure*}

\begin{figure*}[t]
\includegraphics[width=0.86\linewidth]{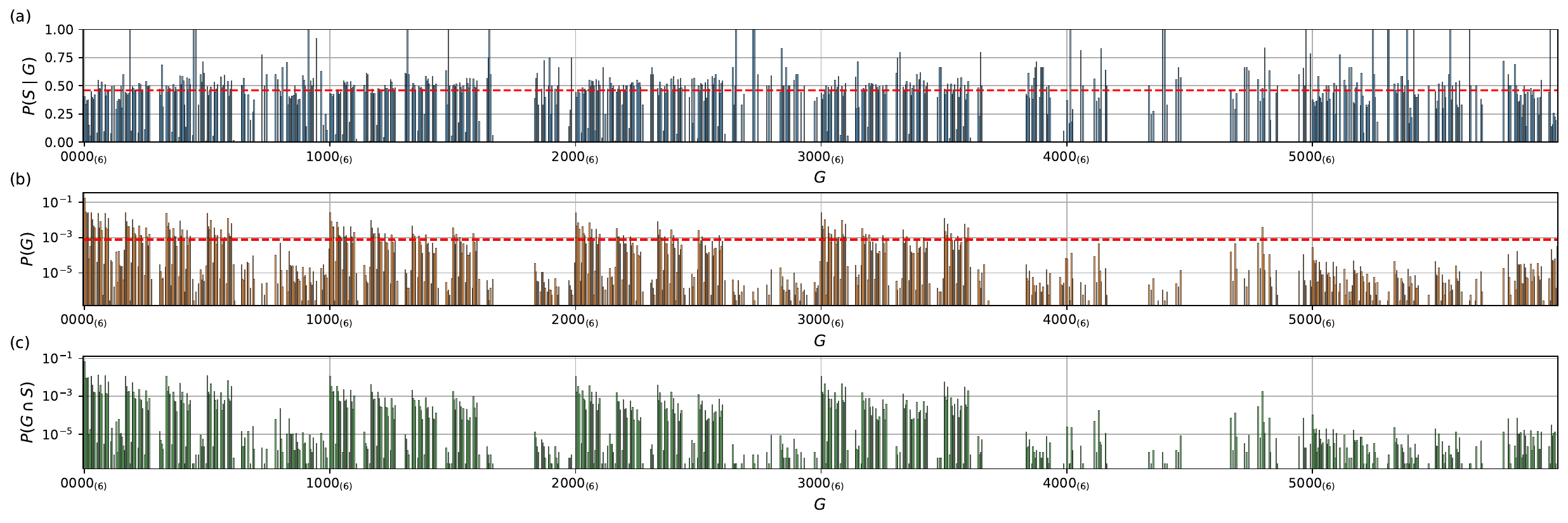}
\caption{\textbf{Pattern-level metrics of pitch sequences with pitch-pattern motif length 4.} (a) Strike probability $P(S|G)$ for a given pitch group $G$, with the expected value indicated by the red dashed line. (b) Probability $P(G)$ of pitch group $G$ showing the relative frequency of each pattern among all pitch sequences (red dashed line: expected value). (c) Conditional probability $P(G \cap S)$ that the pitch pattern $G$ is assessed as the strike. Each pattern corresponds to a motif of length 4, representing four consecutive pitches. All statistics use the qualified dataset containing 3,542,338 pitches.}
\label{fig:pattern_4}
\end{figure*}

\begin{figure*}[t]
\includegraphics[width=0.86\linewidth]{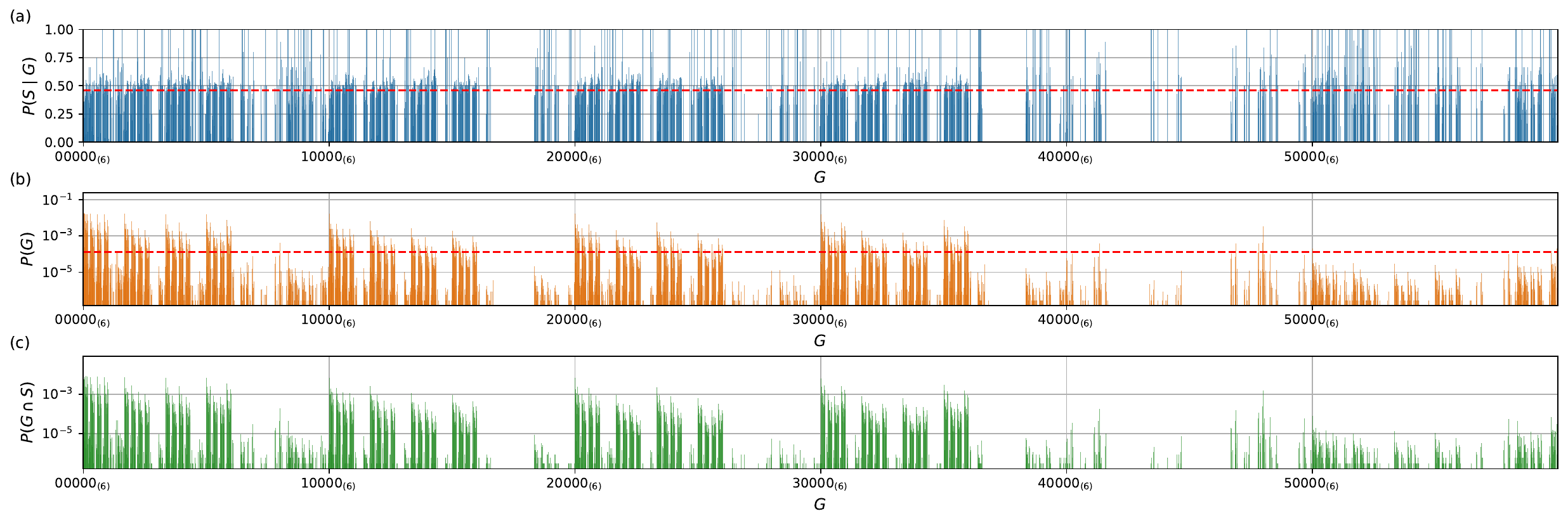}
\caption{\textbf{Pattern-level metrics of pitch sequences with pitch-pattern motif length 5.} (a) Strike probability $P(S|G)$ for a given pitch group $G$, with the expected value indicated by the red dashed line. (b) Probability $P(G)$ of pitch group $G$ showing the relative frequency of each pattern among all pitch sequences (red dashed line: expected value). (c) Conditional probability $P(G \cap S)$ that the pitch pattern $G$ is assessed as the strike. Each pattern corresponds to a motif of length 5, representing four consecutive pitches. All statistics use the qualified dataset containing 3,542,338 pitches.}
\label{fig:pattern_5}
\end{figure*}

These diagnostics allow us to examine how ptich-pattern motif-level statistics evolve as the motif length increases. Consistent with the findings in the main text, we do not observe motif lengths that exhibit a clear or systematic association with win-related outcomes. Figures~\ref{fig:pattern_1}--\ref{fig:pattern_5} present the results sequentially, enabling readers to visually inspect how increasing motif complexity reshapes the distribution and structure of pitch-pattern usage.

\end{document}